\documentclass[11pt]{article}
\usepackage[totalwidth=469pt,totalheight=640pt]{geometry}
\usepackage{amsmath,amssymb,mathrsfs}
\usepackage{psfrag,cite}

\newcommand{\fract}[2]{{\textstyle\frac{#1}{#2}}}
\newcommand\RR{{\mathbb R}}
\newcommand{\ri}{\right}
\newcommand{\lf}{\left}
\newcommand\eq{\begin{equation}}
\newcommand\en{\end{equation}}
\newcommand\bea{\begin{eqnarray}}
\newcommand\eea{\end{eqnarray}}
\newcommand\nn{\nonumber}
\newcommand{\resection}[1]{\setcounter{equation}{0}\section{#1}}

\newcommand{\bg}{{\bf g}}
\begin{document}
\begin{titlepage}
\vskip 0.5cm
\vskip .7cm
\begin{center}
{\Large{\bf Bethe Ansatz equations for the classical 
 $A^{(1)}_{n}$  affine Toda
    field theories
}}
\end{center}
\vskip 0.8cm \centerline{Panagiota Adamopoulou and Clare Dunning} %
\vskip 0.3cm \centerline{\sl\small SMSAS, University of
Kent, Canterbury, CT2 7NF, United Kingdom}

\vskip 0.2cm \centerline{E-mails:}
\centerline{pma7@kent.ac.uk, t.c.dunning@kent.ac.uk}
\vskip 1.25cm
\begin{abstract}
\noindent
We establish a correspondence between classical $A_{n}^{(1)}$ affine Toda field
theories and $A_{n}$ Bethe Ansatz
systems. 
We show that the connection coefficients relating specific
solutions of the associated classical linear problem satisfy functional
relations of the type that appear in the context of the massive
quantum integrable model. 
\end{abstract}

\bigskip

{\small


 }

\end{titlepage}
\setcounter{footnote}{0}
\def\thefootnote{\fnsymbol{footnote}}
%
\resection{Introduction}
\label{intro}
In \cite{Anharmonic_DT}, a connection was established 
between a linear ordinary differential equation (ODE) defined in the
complex plane, and the conformal field theory limit of a certain
quantum integrable lattice model (IM). 
The spectral
determinants of  the ODE introduced in \cite{Anharmonic_DT} satisfy 
Bethe ansatz equations of $A_1$-type, which are objects well known in
the study of quantum integrable systems. 
 Moreover the spectral 
determinants and the Stokes multipliers of the ODE are related by precisely
the same 
functional relations as  the vacuum eigenvalues of the $Q$-operators and
transfer matrix operators of the relevant quantum integrable model
\cite{Anharmonic_DT,BLZ_98,Suzuki:1999rj, TQ_DT}.

Various instances of the correspondence are now known. Of particular
relevance is  
the extension to $SU(n)$ 
Bethe ansatz systems \cite{SU_3,Suzuki_A_n,SU_n}.
Bethe ansatz systems corresponding to simple Lie algebras of
$ABCD$-type have been related to certain pseudo-differential 
equations~\cite{Pseudo-diff}.

Until recently, this ODE/IM correspondence related 
the spectrum of the relevant  ODEs to the Bethe systems of
massless quantum field theories. However, work in 
supersymmetric gauge theory  
\cite{Gaiotto:2009hg,Alday:2009dv} gave a clue as to how the 
 massive quantum field theories may be brought into the correspondence. 
Subsequently, 
Lukyanov and Zamolodchikov \cite{SinhG_Z_L}  presented a way
to obtain the vacuum eigenvalues of the $T$- and $Q$-operators  
of the massive quantum sine-Gordon model starting from
an integrable partial differential equation  related to
 the well-known
classical sinh-Gordon model.  The correspondence is also known  
 for the Bullough-Dodd model
or $A_{2}^{(2)}$ Toda field theory \cite{BD_Dorey_Tateo}.

In this paper we establish the correspondence 
for Bethe ansatz systems  associated with   massive quantum field
theories of type  
$A_{n{-}1}$, starting from the classical
$A^{(1)}_{n{-}1}$ affine Toda field theories. In \S2 we describe a
set of partial differential equations obtained from the affine Toda
field theories by a change of independent and dependent variables, and 
give the associated Lax pairs and linear problems.
The explicit details of the correspondence for
the $A_{2}$-model are presented in \S3, the results for the   
 general $A_{n}$ case are summarised in \S4 and the conclusions are
 found in \S5.

\resection{$A^{(1)}_{n{-}1}$ Toda field theory}

The two-dimensional $A_{n{-}1}^{(1)}$ Toda field
theories are described by the Lagrangian \cite{2dToda_Mikh,Toda_MOP}
\eq
{\cal L} = \frac{1}{2} \sum _{i=1}^{n} (\partial_{t} \eta_{i} ) ^{2} -
( \partial_{x} \eta_{i} )^{2} - \sum_{i=1}^{n} \exp ( 2\eta_{i+1}-
2\eta_{i} )
\label{An_Lagrangian}
\en
with $\eta_{n+1}(x,t)=\eta_1(x,t)$ and 
$\sum_{i=1}^{n} \eta_{i} =0\,.$ 
In light-cone coordinates, 
$
w=x{+}t\,,   \bar{w}=x{-}t\,,$
the corresponding equations of motion are 
\begin{equation}\label{Toda eqns An lc}
2 \,\partial_{\bar{w}}\partial_{w}\eta_{i} = \exp (2 \eta_{i} - 2 \eta_{i-1} )  - \exp ( 2 \eta_{i+1} - 2 \eta_{i})  \quad \mbox{with} \quad i=1, \ldots, n\,.
\end{equation}
However, these are not the partial differential equations through which
we will establish a connection to Bethe ansatz systems of
$A_{n{-}1}$-type. For the simplest case ($n=2$) the relevant equation
is found by modifiying the 
sinh-Gordon equation  using the changes of variables 
$dw=p(z)^{1/2} \,dz$ and $d{\bar w}=p(\bar z)^{1/2}\,d{\bar z} $
where 
$p(z)=z^{2M} - s^{2M}$ \cite{SinhG_Z_L}. Sending  $\eta=\eta-
\log (p(z) p(\bar z)/4$, the sinh-Gordon equation becomes
\eq
\partial_z \partial_{\bar z} - \exp(2\eta(z, \bar z)) + p(z) p(\bar z)
\,\exp(-2\eta(z,\bar z)) =0~.
\label{sG}
\en
The introduction of the function $p(z)$ brings two parameters $M$ and $s$,
which are related to the coupling and the mass scale of the quantum
sine-Gordon model. 
For general $n$, motivated by the examples involving a single Toda field \cite{SinhG_Z_L, BD_Dorey_Tateo}, 
we make the  change of variables 
\begin{equation}\label{change_of_var w_z An}
{\rm d}w = p(z)^\frac{1}{n} \,{{\rm d}z}
\,, \quad  {\rm d}{ \bar w}   = p({\bar z})^\frac{1}{n}
\,{{\rm d}{\bar z}}
\end{equation} 
and introduce the  function $p(t)=t^{nM}{-}s^{nM}$ 
where the real, positive parameter $s$ is related to the mass scale in
the associated quantum model and 
$M$ is related to the coupling. 
Setting 
\eq \label{shift}
\eta_{i} (z,\bar z)\rightarrow \eta_{i} (z,\bar z) +
\frac{n-(2i-1)}{4n}\ln(p(z)p(\bar{z}))\,, 
\en
the modified affine Toda field equations are 
\bea
2\,\partial_{\bar{z}}\partial_{z}\eta_{1} &=&
p(z)p(\bar{z})\,\mbox{e}^{2\eta_{1} - 2\eta_{n} }  -
\mbox{e}^{2\eta_{2} - 2\eta_{1}}  \,, \nn \\
2\,\partial_{\bar{z}}\partial_{z}\eta_{i} &=& \mbox{e}^{2\eta_{i} -
  2\eta_{i-1} }  - \mbox{e}^{2\eta_{i+1} - 2\eta_{i}}  \quad
\mbox{for} \quad i=2, \ldots, n-1\,, \label{Generalised Toda An} \\
2\,\partial_{\bar{z}}\partial_{z}\eta_{n} &=& \mbox{e}^{2\eta_{n-1} -
  2\eta_{n} }  - p(z)p(\bar{z})\,\mbox{e}^{2\eta_{1} - 2\eta_{n}}
\,.\nn
\eea 

The equations  \eqref{Generalised Toda An} can alternatively be viewed as
arising 
from the zero-curvature condition $V_{z} - U_{\bar z} + [U,V]=0$ 
of 
the linear problem 
\begin{equation}\label{linear prob generalised An}
\big( \partial_{z} + U( z,\bar{z},\lambda) \big) \mathbf{\Psi}=0\,, \quad \big( \partial_{\bar{z}} + V(z,\bar{z},\lambda) \big) \mathbf{\Psi}=0\,,
\end{equation}
where
$\lambda=\exp(\theta)$ is the   spectral parameter  and 
\begin{equation}\label{U generalised An}
(U(z,\bar{z}, \lambda))_{ij}=\partial_{z} \eta_{i} \,\delta_{ij} + \lambda\,
(C(z))_{ij}\,, \quad (V(z,\bar{z},\lambda))_{ij} = -\partial_{\bar{z}} \,\eta_{i}\,
\delta_{ij} + \lambda^{-1} (C(\bar z))_{ji}\,,
\end{equation}
\begin{equation}\label{C1 of U generalised An}
(C(z))_{ij} = \left\{ \begin{array}{cll} 
 \exp(\eta_{j+1}-\eta_{j}) \,\delta _{i-1,j} &, & j=1, \ldots, n{-}1 \\
p(z)\,\exp(\eta_{j+1}-\eta_{j}) \,  \delta _{i-1,j} & , & j=n\,,
\end{array}\right.
\end{equation}
where 
\eq
\delta_{ij} = \begin{cases}
1 & {\rm if \ }i \equiv j \mod n \\ 
0 & {\rm otherwise}\,.
\end{cases}
\en

The Lax pair \eqref{linear prob generalised An} for the
modified Toda field equations \eqref{Generalised Toda An} is related
to that associated with the standard $A_{n-1}^{(1)}$ Toda
equations. We start from the Lax pair  for the
 $A_{n-1}^{(1)}$  Toda equations presented in \cite{2dToda_Mikh},
written in light-cone coordinates $(w,\bar{w})$ as 
\eq \label{Ubar0,Vbar0}
\big( \partial_{w} + \widehat{U}( w,\bar{w},\lambda) \big) \mathbf{\Phi}=0\,, \quad \big( \partial_{\bar{w}} + \widehat{V}(w,\bar{w},\lambda) \big) \mathbf{\Phi}=0\,
\en
noting that the zero-curvature condition $\widehat V_{w} - \widehat U_{\bar w} +
[\widehat U,\widehat V]=0$ is equivalent to \eqref{Toda eqns An lc}. 
 We apply the change
of variables  \eqref{change_of_var w_z An} and the transformation
\eqref{shift} to \eqref{Ubar0,Vbar0} to obtain 
\eq \label{Ubar,Vbar}
\big( \partial_{z} + \widetilde{U}( z,\bar{z},\lambda) \big)
\mathbf{\Phi}=0\,, \quad \big( \partial_{\bar{z}} +  \widetilde{V}(z,\bar{z},\lambda) \big) \mathbf{\Phi}=0
\en
where 
\eq
\widetilde U(z,\bar z ,\lambda)=\widehat U(w \to z,\bar w \to \bar z ,\lambda)
\en
and the shift of the Toda fields \eqref{shift} in $\widehat U$ is implied.
The  linear problem \eqref{Ubar,Vbar} is related to the
modified linear problem \eqref{linear prob generalised An} by the gauge transformation
\bea
 U(z,\bar z,\lambda)&=& g^{-1} \,g_z +  g^{-1} \,\widetilde U(z,\bar{z},\lambda) \,g \,, \\ 
V(z,\bar z,\lambda)&=& g^{-1} \,g_{\bar{z}} +  g^{-1} \,\widetilde V(z,\bar{z},\lambda) \,g \,,
\eea
with $\mathbf{\Phi}=g \mathbf{\Psi}$ and the matrix $g$ has entries 
\eq
(g)_{ij} = \left(\frac{ p(\bar z)}{p(z)} \right)^{n-\frac{2i-1}{4n}} \,\delta_{ij}~.
\en 
We observe that the function $p(t)$ appears  
in the entries of the Lax 
matrices $U$ and $V$  that are  related to the generator associated to the affine root
of the $A_{n-1}^{(1)}$  Lie algebra.

Since we will be concerned with specific solutions to the
modified version of the $A_{n-1}^{(1)}$ Toda equations that are real-valued,
it is convenient to introduce polar coordinates
$
z=\rho \, \mbox{e}^{i \phi}\,, \bar{z}= \rho \, \mbox{e}^{-i \phi}$
 with $\rho, \phi \in \RR$. However, $z, \bar{z}$ will sometimes be
 treated as independent complex variables.

The modified Toda equations of motion \eqref{Generalised Toda An} are invariant under the discrete symmetry 
\begin{equation}\label{symmetry z zbar An}
z \rightarrow \exp \left( 2 \pi i/nM \right) z \,, \quad \bar{z} \rightarrow \exp \left( -2 \pi i/nM\right)\bar{z} \,.
\end{equation}
The linear problem is  invariant under \eqref{symmetry z zbar An} 
if the spectral parameter is shifted  as 
$\lambda \rightarrow \sigma^{-1/M} \lambda $ 
(or equivalently $\theta \rightarrow \theta {-} 2 \pi i/nM$) where 
 $\sigma=\exp(2
\pi i/n)\,.$ 
It will be useful to define the transformations
\begin{equation}\label{Omega transf An}
\widehat{\Omega}: \quad   \phi\rightarrow \phi +\frac{2\pi}{nM}\,, \quad \theta \rightarrow \theta - \frac{2\pi i}{nM}\,,
\end{equation}
and, for any $n  \times n$ matrix $A(\lambda)$, 
\begin{equation}\label{S transf An}
\widehat{S}: \quad A(\lambda) \rightarrow S\, A(\sigma^{-1}\lambda)
\,S^{-1} \quad  {\rm where\ }\quad 
(S)_{jk} = \sigma^j \,\delta_{jk}\,.
\end{equation}
Such groups of transformations (acting on a loop
algebra) are known as reduction groups \cite{Toda_MOP, Reduction_Mikhailov}. Since $\widehat{S}^{n}=id$ the 
 group generated by $\widehat{S}$ is isomorphic to
$\mathbb{Z}_{n}$. 
The matrices $U$ and $V$ 
are invariant under
the transformations $\widehat{\Omega}$ and $\widehat{S}$:
\bea
\hspace{-.9cm} U\left(\rho, \phi {+} \fract{2 \pi}{nM}, \theta {-}\fract{2 \pi
    i}{nM}\right) = U\left(\rho, \phi, \theta \right)  &,& V\left(\rho, \phi {+} \fract{2 \pi}{nM}, \theta {-}\fract{2 \pi i}{nM}\right) = V\left(\rho, \phi, \theta \right)\,,\\
 \hspace{-.9cm} S\, U\left(\rho, \phi, \theta {-}\fract{2 \pi i}{n}  \right) S^{-1}  =
 U(\rho,\phi,\theta) &,& S\, V\left(\rho, \phi, \theta {-} \fract{2 \pi i}{n} \right) S^{-1}  = V(\rho,\phi,\theta) \,.
\eea 
Applying $\widehat S$ to \eqref{linear prob generalised An} we find 
$S\,\mathbf{\Psi}(\sigma^{-1} \lambda )$ 
satisfies the linear problem, and for each solution $\mathbf{\Psi}$ there
is a constant $c$ such that 
\begin{equation}\label{S Psi = C Psi}
S \,\mathbf{\Psi}(\sigma^{-1} \lambda ) = \sigma^c\,\mathbf{\Psi}(\lambda) \,.
\end{equation}
We will introduce a family of solutions to the linear problem
which respects these symmetries. 

\resection{The $A_{2}$ case \label{sec: Generalised A_2}} 
We  start by explaining the approach for the simplest model 
involving two fields:  
 the $A^{(1)}_{2}$ affine Toda field theory. 
Motivated by \cite{SinhG_Z_L,BD_Dorey_Tateo}, we specify a
two-parameter family of solutions 
to the equations \eqref{Generalised Toda An} 
that respects the discrete symmetry  \eqref{symmetry z zbar An}.
These solutions 
are unique, real and  
 finite everywhere except at $\rho=0$ and 
have periodicity  
$\eta_{i}\big(\rho, \phi+ 2\pi/3M\big)= \eta_{i}(\rho,
\phi)\,.$ The solutions are further specified in terms of the parameters
$g_0,g_2$ via 
their asymptotic behaviour at small and large
values of $\rho$ as 
\bea 
\label{small rho A2}
\eta_{1}(\rho,\phi) = (2-g_2) \ln \rho + O(1) &,&
\eta_{3}(\rho,\phi) = -g_0 \ln \rho + O(1)  \quad \,\mbox{as} \quad \rho
\rightarrow 0\,, \\ 
\label{large rho A2}
\eta_{1}(\rho, \phi) = -M \ln \rho +o(1) &,&    \eta_{3}(\rho,\phi) =
M \ln \rho +o(1) \quad \quad \mbox{as} \quad \rho \rightarrow \infty~.
\eea 
The  coefficients have been chosen to ensure consistency
with the notation in 
the massless limit \cite{SU_3}.

The small-$\rho$ asymptotic of $\eta_i(\rho,\phi)$  may be improved by
noting that the relevant solutions $\eta_i(w ,\bar w)$ to the
standard Toda field 
equations \eqref{Toda eqns An lc} depend only on $w\bar w$.
Under the symmetry reduction 
$t = (2w\bar{w})^{1/2}\,,$ 
the corresponding equations of motion become a coupled system  of
ODEs  for $\eta_{i}(w,\bar{w}) = y_{i}(t):$
\begin{equation}\label{Toda eqns reduction A2}
\begin{array}{l}
{\displaystyle   \frac{d^{2}}{dt^{2}}\, y_{1} + \frac{1}{t} \frac{d}{dt}\, y_{1} + \mbox{e}^{-4y_{1} -2y_{3}} - \mbox{e}^{2y_{1}-2y_{3}} =0}\,,\\
\\
{\displaystyle    \frac{d^{2}}{dt^{2}}\, y_{3} + \frac{1}{t} \frac{d}{dt}\, y_{3} + \mbox{e}^{2y_{1} -2y_{3}} - \mbox{e}^{4y_{3}+2y_{1}} =0}\,.
\end{array}
\end{equation}

To determine the asymptotic behaviour  of \eqref{Toda eqns reduction
  A2}
we use the method of
dominant balance (see, for example, \cite{Asymptotics_White}).
The first two terms of  each equation in \eqref{Toda eqns
  reduction A2} balance  in the limit  $t \rightarrow 0$  if 
\begin{equation}\label{leading order asymt 0 A2}
y_{1} (t) \sim (2-g_2) \ln t + b_{1}\,, \quad y_{3}(t) \sim -g_0 \ln t + b_{3}\,,
\end{equation}
where  $b_i$ are
arbitrary constants  and for dominance the real constants
$g_i$  must satisfy  
\begin{equation}\label{relation l1 l3}
g_0<g_1<g_2 \quad,\quad 2g_0+g_1>0\,. 
\end{equation}  
The additional parameter $g_1$ appears in the 
asymptotic of the eliminated field $\eta_2=y_2$ (recall
$\eta_1+\eta_2+\eta_3=0$) and is such that  $g_0+g_1+g_2=3\,.$

Extending the analysis, we find that the
sub-leading contributions to \eqref{leading order
  asymt 0 A2} take the 
form of two power series in particular powers of $t$ \cite{pma}.  
Using these results, we deduce the asymptotic behaviours as $\rho\to 0$ 
   of the required solutions of the modified Toda equations 
\eqref{Generalised Toda An} 
are 
\begin{eqnarray}\label{full asymptotic generalised sol1 A2}
\eta_{1} (z,\bar{z}) &\sim & (1-g_2/2) \ln (z\bar{z}) + b_{1} +
\sum_{k=1}^{\infty} 
 \frac{(-1)^{2k+1}}{6k \, s^{3kM}}
 \left( z^{3kM} + \bar{z}^{3kM} \right)  
\nonumber\\
 &+& \sum_{m=1}^{\infty} c_{m}\, (z\bar{z})^{m \left( g_0-g_2+3 \right)} -  \sum_{m=1}^{\infty} d_{m}\, (z\bar{z})^{m \left( g_0+2g_2-3 \right)}\,, 
\end{eqnarray}
and 
\begin{eqnarray}\label{full asymptotic generalised sol3 A2}
\eta_{3} (z,\bar{z}) & \sim &  -(g_0/2) \ln (z\bar{z}) + b_{3}  -
\sum_{k=1}^{\infty} 
 \frac{(-1)^{2k+1}}{6k \,s^{3kM}}
\left( z^{3kM} + \bar{z}^{3kM} \right) 
\nonumber\\
&+& \sum_{m=1}^{\infty} e_{m}\, (z\bar{z})^{m \left( 3-2g_0-g_2 \right)} -  \sum_{m=1}^{\infty} f_{m}\, (z\bar{z})^{m \left( g_0-g_2+3 \right)}\,. 
\end{eqnarray}
The series in single powers of $z,\bar z$ arises from the functions
$p(z), p(\bar z)\,.$ We have redefined the constants $b_i$ to
incorporate all constant corrections, and  
the coefficients $c_{m},d_{m},e_{m}$ and  $f_{m}$ may be
determined recursively.
For
example
\eq
c_1=f_1=\frac{s^{6M} \mbox{e}^{2b_{1} - 2b_{3}}}{ 2(3{+}g_0{-}g_2)^{2}} \
, \  
d_1=\frac{\mbox{e}^{-4b_{1} - 2b_{3}}  }{2(g_0{+}2g_2{-}3)^{2}} \ ,  \ 
e_1=\frac{ \mbox{e}^{4b_{3} + 2b_{1}}}{2 (3{-}2g_0{-}g_2)^{2}}\,.
\en

The leading logarithmic contribution to the large-$\rho$ expansion of
\eqref{large rho A2}  arises from the term  $p(z)p(\bar z)$ because,
arguing as above,   in the large-$t$ limit we obtain  
\begin{equation} \label{y1,3 order 1}
y_{1}(t) = O(1)\,, \quad y_{3}(t) = O(1) \quad \mbox{as} \quad t \rightarrow \infty \,.
\end{equation}

From the linear system \eqref{linear prob generalised An} 
we may obtain a pair of third order linear ODEs for two of the components of 
$\mathbf{\Psi}= \left( \Psi_{1},\Psi_{2}, \Psi_{3}
\right)^{\mathrm{T}}$.  
Setting $\Psi_{3} =
\exp(\eta_{3})\, \psi$ and $\Psi_{1}= \exp(-\eta_{1})\,
\bar{\psi}$, we define a 
 general solution to the linear system through 
\begin{eqnarray}\label{Psi A2}
\mathbf{\Psi}(z,\bar{z},\lambda) &=& 
 \left(\begin{array}{c}
\lambda^{-2} \, \mbox{e}^{3\eta_{1}+2\eta_{3}} \, \partial_{z}\left( \mbox{e}^{-2\eta_{1} - 4\eta_{3} } \, \partial_{z} \left(  \mbox{e}^{2\eta_{3}}\, \psi \right) \right)  \\
-\lambda^{-1} \, \mbox{e}^{-\eta_{1}-3\eta_{3}}\,  \partial_{z} \left(  \mbox{e}^{2\eta_{3}}\, \psi \right)   \\
 \mbox{e}^{\eta_{3}}\psi
 \end{array}\right) \nonumber\\
&=&
 \left(\begin{array}{c}
\mbox{e}^{-\eta_{1}} \, \bar{\psi}   \\
-\lambda \, \mbox{e}^{3\eta_{1} + \eta_{3}} \, \partial_{\bar{z}} \left( \mbox{e}^{-2\eta_{1}}\, \bar{\psi} \right)   \\
\lambda^{2} \, \mbox{e}^{-2\eta_{1} - 3\eta_{3}} \, \partial_{\bar{z}} \left( \mbox{e}^{4\eta_{1}+2\eta_{3}} \, \partial_{\bar{z}} \left( \mbox{e}^{-2\eta_{1}}\, \bar{\psi} \right)  \right)
 \end{array}\right)\,.
\end{eqnarray} 
Then applying $\partial_z+U$ and $\partial_{\bar z} +V$  to $\mathbf{\Psi}$
  we deduce $\psi$ and $\bar{\psi}$ satisfy third order ODES: 
\begin{equation}\label{ODE psi A2}
\partial_{z}^{3}\psi + u_{1}(z,\bar{z})\, \partial_{z} \psi +  \left( u_{0}(z,\bar{z}) + \lambda^{3}p(z) \right) \psi=0\,,
\end{equation}
\begin{equation}\label{ODE psibar A2}
\partial_{\bar{z}}^{3} \bar{\psi} + \bar{u}_{1}(z,\bar{z})\, \partial_{\bar{z}} \bar{\psi} +  \left( \bar{u}_{0}(z,\bar{z}) + \lambda^{-3}p(\bar{z}) \right) \bar{\psi}=0\,,
\end{equation}
with
\begin{equation}\label{u0, u1 A2}
\begin{array}{l}
u_{1}(z,\bar{z}) = -2 \left( 2 \left(\partial_{z}\eta_{1}\right)^{2} + 2 \partial_{z}\eta_{1} \partial_{z}\eta_{3} + 2\left(\partial_{z}\eta_{3}\right)^{2} +  \partial_{z}^{2} \eta_{1} -  \partial_{z}^{2}\eta_{3}   \right)\,, \\
u_{0}(z,\bar{z})= -4\partial_{z}\eta_{3} \big( 2\partial_{z}\eta_{1} \partial_{z}(\eta_{1}+\eta_{3}) +\partial_{z}^{2}\eta_{1} + 2\partial_{z}^{2}\eta_{3} \big) +2\partial_{z}^{3}\eta_{3} \,,
\end{array}
\end{equation}
and 
\begin{equation}\label{u0bar, u1bar A2}
\begin{array}{l}
\bar{u}_{1}(z,\bar{z}) = -2 \left( 2 \left(\partial_{\bar{z}}\eta_{1}\right)^{2} + 2 \partial_{\bar{z}}\eta_{1} \partial_{\bar{z}}\eta_{3} + 2\left(\partial_{\bar{z}}\eta_{3}\right)^{2} +  \partial_{\bar{z}}^{2} \eta_{1} -  \partial_{\bar{z}}^{2}\eta_{3}   \right)\,, \\
\bar{u}_{0}(z,\bar{z})= 4\partial_{\bar{z}}\eta_{1} \big( 2\partial_{\bar{z}}\eta_{3} \partial_{\bar{z}}(\eta_{1}+\eta_{3}) -\partial_{\bar{z}}^{2}\eta_{1} - \partial_{\bar{z}}^{2}\eta_{3} \big) -2\partial_{\bar{z}}^{3}\eta_{1} \,.
\end{array}
\end{equation}

\subsection{The $Q$ functions}

We now find the behaviour of ${\bf \Psi}$ for small and large
values of $\rho$.  Setting $\psi=z^\mu$ and treating $\bar z$ as a
fixed parameter as $z\to 0$,  the roots of the indicial polynomial of \eqref{ODE
    psi A2}  in this limit 
 provide three different solutions, defined by 
\begin{equation}\label{small rho solutions A2}
\chi _{0}\sim z^{g _{0}}, \quad \chi _{1} \sim z^{g _{1}}, \quad
\chi_{2} \sim z^{g _{2}}\,,\quad \quad z \rightarrow 0.
\end{equation}
Consequently the linear problem \eqref{linear prob generalised An}  has  solutions for small
$\rho$ defined by
the leading order behaviours as follows:
\begin{equation}\label{Psi small rho A2}
\mathbf{\Xi}_{0}  \sim \left(\begin{array}{c} 
0\\
0\\
\mathrm{e}^{g_{0}(\theta +i\phi)}\\
\end{array}\right)\,,
\ 
\mathbf{\Xi}_{1}  \sim \left(\begin{array}{c} 
0\\
\mathrm{e}^{(g_{1}-1)(\theta +i\phi)}\\
0\\
\end{array}\right)\,, 
\
\mathbf{\Xi}_{2} \sim \left(\begin{array}{c} 
\mathrm{e}^{(g_{2}-2)(\theta +i\phi)}\\
0\\
0\\
\end{array}\right)\,
\end{equation}
where $\mathbf{\Xi}_{j} \equiv \mathbf{\Xi}_{j} (\rho,
\phi, \theta, \mathbf{g} )$ and $\mathbf{g} = \{g_{0}, g_{1}, g_{2}
\}.$ 
The $\theta$-dependent constants have been introduced in
 \eqref{Psi small rho A2} to ensure the solutions are 
invariant under the symmetry $\widehat{\Omega}$ \eqref{Omega transf
  An}. The matrix of the solutions is normalised as  
$
\det \left(  \mathbf{\Xi}_{0}, \mathbf{\Xi}_{1}, \mathbf{\Xi}_{2}
\right) = -1 \,.$ 

Concentrating instead on  the large-$\rho$ limit, 
the ODEs  \eqref{ODE psi
  A2} and \eqref{ODE psibar A2}  reduce  to
\begin{equation}\label{ODE psi large rho A2}
\partial_{z}^{3} \psi + \lambda^{3}p(z)\, \psi=0
\quad,\quad \partial_{\bar{z}}^{3} \bar{\psi} + \lambda^{-3}p(\bar{z})\, \bar{\psi}=0\,.
\end{equation}
Using a WKB-type of analysis (see  \cite{SU_3}), 
 there exist solutions    with  asymptotic
behaviour  for $M>1/2$ 
\begin{equation}
\psi \sim  z^{-M} \exp \left( -\lambda\, \frac{z^{M+1}}{M+1}
  +f_{1}(\bar{z}) \right) 
\quad ,\quad 
\bar{\psi} \sim  \bar{z}^{-M} \exp \left( -\lambda^{-1} \frac{\bar{z}^{M+1}}{M+1} +f_{2}(z) \right)
\end{equation}
as $\rho \rightarrow \infty$ with $|\phi| <4 \pi/(3M{+}3)$  where
$z=\rho \,\exp (i\phi)$. In this sector of the complex plane, 
these solutions  are the unique
solutions for $\lambda \in \RR$  that tend to zero fastest
as $\rho \to 
\infty$. The functions $f_{1}(\bar{z})$ and $f_{2}(z)$ are fixed by
equating any line of  \eqref{Psi A2} in the limit $\rho
\to \infty$:  
\begin{equation}
 f_{1}(\bar{z})= -\lambda^{-1}\frac{\bar{z}^{M+1}}{M+1}\,, \quad f_{2}(z)= -\lambda\frac{z^{M+1}}{M+1}\,.
\end{equation}    
Hence the large-$\rho$ asymptotic solution $\mathbf{\Psi} $ to the linear problem \eqref{linear prob generalised An} is 
\begin{equation}\label{Psi WKB A2}
\mathbf{\Psi} \sim  \left(\begin{array}{c} 
\mathrm{e}^{i\phi M}\\
1\\
\mathrm{e}^{- i\phi M}\\
\end{array}\right) \exp \left( -2 \, \frac{\rho^{M+1}}{M+1} \cosh (\theta + i\phi(M+1)) \right) \quad \mbox{as} \quad \rho \rightarrow \infty\,.
\end{equation}

Since the solutions $\mathbf{\Xi}_{j}$ 
defined  at small-$\rho$  by \eqref{Psi small rho
  A2} are linearly independent, we  take $\{{\bf \Xi}_j\}$ as a basis for the
space of all solutions to the linear problem.
 Hence we can express $\mathbf{\Psi}$ in this basis as 
\begin{equation}\label{Xi Psi connection A2}
\mathbf{\Psi} = Q_{0}(\theta,\mathbf{g})\, \mathbf{\Xi}_{0} +  Q_{1}(\theta,\mathbf{g}) \, \mathbf{\Xi}_{1} +  Q_{2}(\theta,\mathbf{g}) \,\mathbf{\Xi}_{2}\,,
\end{equation} 
where $\mathbf{\Psi}$ and 
$\mathbf{\Xi}_{j}$ depend implicitly on $(\rho, \phi, \theta,
\mathbf{g})$ and 
\begin{equation}\label{Qi as determinats}
\hspace{-0.11cm} Q_{0} = {-}\det \left( \mathbf{\Psi},
  \mathbf{\Xi}_{1}, \mathbf{\Xi}_{2} \right) \,,  \ 
Q_{1} = {-}\det \left( \mathbf{\Xi}_{0}, \mathbf{\Psi},
  \mathbf{\Xi}_{2} \right) \,,  \ 
Q_{2} = {-}\det \left( \mathbf{\Xi}_{0}, \mathbf{\Xi}_{1}, \mathbf{\Psi} \right). 
\end{equation}

We note that the functions $Q_j(\theta,{\bf g})$  are quasi-periodic
functions of $\theta$.   We use the small-$\rho$ asymptotics \eqref{Psi small rho A2} to determine  the constants $c_j$ such
that the relation  $S \,\mathbf{\Xi}_{j}(\sigma^{-1} \lambda) =c_j
\,\mathbf{\Xi}_{j}(\lambda)$ holds, then act with the symmetry $
\widehat{\Omega}$ to obtain  
\eq \label{Xi new}
S \, \mathbf{\Xi}_{j} \left( \rho, \phi + \fract{2\pi}{3M}, \theta
  -\fract{2 \pi i}{3M} - \fract{2 \pi i}{3} \right) =\exp( -g_{j}
\fract{2 \pi i}{3}) \, \mathbf{\Xi}_{j} (\rho, \phi , \theta)  \,.
\end{equation}
It is also straightforward to prove that 
\begin{equation}\label{combined tranf invariance Psi}
S \, \mathbf{\Psi} \left( \rho, \phi + \fract{2 \pi}{3 M} , \theta -
  \fract{2 \pi i}{3 M} - \fract{2 \pi i}{3} \right) = \exp \lf(\fract{4
  \pi i}{3}) \, \mathbf{\Psi} (\rho, \phi , \theta \ri)\,. 
\end{equation}
Inserting  \eqref{Xi
new} and  
\eqref{combined tranf invariance Psi} into the expressions for $Q_j$
given in \eqref{Qi as
  determinats}, and  recalling that $S^n=\mathbb{I}$, we obtain  
\begin{equation}
Q_{j} (\theta,{\bf g}) = \exp\lf( -\fract{2 \pi i}{3}(g_{j}-1)\ri) \, Q_{j} \big( \theta - \fract{2 \pi i}{3}\fract{(M+1)}{M} ,{\bf g}\big)\,, \quad \mbox{with} \quad j=0,1,2\,.
\end{equation}
Before we derive $A_2$-type functional
relations for the  $Q_j$, we show that in a specific limit 
the differential equations obtained from the modified $A_2$ linear system
reduce to those of the massless $SU(3)$ ODE/IM
correspondence \cite{SU_3}.

\subsection{Conformal limit}\label{sec:Conformal limit A2}

We initially focus on the ODE
for $\psi$  \eqref{ODE psi A2}.  We take the parameter  $\bar z $ to zero and use the
asymptotics  \eqref{full asymptotic generalised sol1 A2} and 
\eqref{full asymptotic generalised sol3 A2} 
to simplify the contribution to
\eqref{ODE psi A2} from the Toda fields 
$\eta_i$. If we now set
\begin{equation}\label{x E definition A2}
x=z\, \exp \left(\fract{\theta}{M+1} \right)\,, \quad  E=s^{3M}\exp \left(\fract{3M\theta}{M+1} \right)\,,  
\end{equation}
then as   $z \sim s \to 0$ and $\theta \to
\infty$, the new variables $x,E$ are finite.  As a result, the 
ODE \eqref{ODE psi A2} reduces to
\begin{equation}\label{SU(3) ODE A2}
\left( \partial_{x}^{3} +\frac{\left( g_{0}g_{1} + g_{0}g_{2} + g_{1}g_{2} -2 \right)}{x^{2}}\,  \partial_{x} -\frac{ g_{0}g_{1}g_{2}}{x^{3}} +p(x,E)\right) \psi =0
\end{equation}
with $p(x,E)= x^{3M} - E$, precisely matching   the third-order ODE
introduced in \cite{SU_3} in 
the context of the (massless) ODE/IM correspondence. In this massless
(or conformal) limit the
functions 
 $Q_{j}$ are related to the vacuum eigenvalues of certain
$Q$-operators studied in \cite{W3} in the context of ${\cal W}_3$ conformal field
theory. Therefore, we anticipate the $Q_{j}$ in \eqref{Xi Psi connection A2} are
related to the integrable structures of the massive  $A_{2}$ quantum field
theory model. 

Alternatively, sending the parameter $z \to 0$ 
in the ODE for $\bar{\psi}$ \eqref{ODE
  psibar A2}  while keeping $\bar z$ small but finite, then 
 taking the limit $\bar z \sim s \to 0 $ as  $\theta \rightarrow -\infty$
with
\begin{equation}\label{x bar E bar definition A2}
\bar{x}=\bar{z}\, \exp \left(-\fract{\theta}{(M+1)} \right)\,, \quad \bar{E}=s^{3M} \exp \left(-\fract{3M\theta}{(M+1)} \right)\,,  
\end{equation}
yields 
\begin{equation}\label{ODE psibar A2 cf}
\left(   \partial_{\bar{x}}^{3} +\frac{( g^{\dagger}_{0}
    g^{\dagger}_{1} + g^{\dagger}_{0} g^{\dagger}_{2} +
    g^{\dagger}_{1} g^{\dagger}_{2} -2 )}{\bar{x}^{2}} \, 
\partial_{\bar{x}} -\frac{g^{\dagger}_{0} g^{\dagger}_{1} g^{\dagger}_{2}  }{\bar{x}^{3}}+p(\bar{x}, \bar{E})  \right) \, \bar{\psi} =0\,,
\end{equation}
where $
g_{i}^{\dagger}=2 - g_{2-i}\,.$ Note that \eqref{ODE psibar A2 cf} is
the adjoint equation to  \eqref{SU(3) ODE A2}.

\subsection{Functional relations}\label{sec:functional relations
  A_2}
We now return to the study of \eqref{ODE psi A2} and \eqref{ODE psibar A2}. We establish via a set of functional relations  
 the Bethe ansatz equations matching those of the associated massive quantum integrable
model. 
In terms of the
variables \eqref{x E definition A2}, 
the ODE \eqref{ODE psi A2}  is  
\begin{equation}\label{ODE psi x A2}
\lf( \partial_{x}^{3} + u_{1}(x,\bar{x})\, \partial_{x}  +  \left(
  u_{0}(x,\bar{x}) + p(x,E) \right)  \ri)\psi(x,\bar x,E,\bar E, {\bf g})=0\,,
\end{equation}
where $u_{0}$ and $u_{1}$ are functions of the Toda fields
$\eta_{1}$, $\eta_{3}$. 
We treat $x$ as a complex variable, $\bar x$ as a parameter and  define sectors in the complex
$x$-plane as 
\begin{equation}\label{Stokes sectors A2}
\mathcal{S}_{k}:\left|\arg x-\frac{2k\pi}{3(M+1)}\right|<\frac{\pi}{3(M+1)}\,.
\end{equation}
Then 
\begin{equation}\label{psi k A2}
\psi_{k}(x,\bar{x},E, \bar{E}, \mathbf{g}) = \omega^{k} \psi
(\omega^{-k}x,\omega^{k}\bar{x},\omega^{-3kM} E,\omega ^{3kM} \bar{E},
\mathbf{g} ) \,, \  \ 
\omega = \mbox{e}^{\frac{2\pi i}{3(M+1)}}\,,
\end{equation}
 are all solutions to
\eqref{ODE psi x A2} for  integer $k$. 
These solutions have large-$x$ asymptotics 
\begin{equation}\label{WKB psi k A2}
\psi_{k} \sim \omega^{k(M+1)} \frac{x^{-M}}{i\sqrt{3}} \exp \left( -\omega^{-k(M+1)}\frac{x^{M+1}}{M+1} -\omega^{k(M+1)} \frac{\bar{x}^{M+1}}{M+1} \right)\,
\end{equation}
for $x \in
\mathcal{S}_{k-\frac{3}{2}} \cup \mathcal{S}_{k- \frac{1}{2}} \cup
\mathcal{S}_{k+ \frac{1}{2}} \cup \mathcal{S}_{k+ \frac{3}{2}}.$ 
Moreover,  each $\psi_k$ is the unique solution that decays to zero
fastest for large $x$ within the Stokes sector ${\cal S}_k\,.$
Defining  the notation
\eq 
W_{k_{1},\ldots,k_{m}}=W[\psi_{k_{1}},\ldots,\psi_{k_{m}} ]
\en where
$W$ denotes the Wronskian of $m$ functions, it follows from
the asymptotic \eqref{WKB psi k A2} that the Wronskian
$W_{k,k+1,k+2}=1$. Thus any 
triplet $\{\psi_k,\psi_{k+1},\psi_{k+2}\}$   is a basis of solutions
to \eqref{ODE psi x A2}, and we write 
\begin{equation}\label{psi0 in terms of psi123}
\psi= C^{(1)}(E,\bar{E},\mathbf{g})\, \psi_{1} +
C^{(2)}(E,\bar{E},\mathbf{g})\,\psi_{2} + C^{(3)}(E,\bar{E},{\mathbf
  g})\,\psi_{3}\,, 
\end{equation}
where the Stokes multipliers $C^{(j)}(E,\bar{E},\mathbf{g})$ are
analytic functions and the dependence of $\psi_{k}$ on
$(x,\bar{x},E,\bar{E},\mathbf{g})$ has been omitted. 
Taking the Wronskian of \eqref{psi0 in terms of psi123} with
$\psi_{2}$, using $C^{(3)}=1$  and the  determinant relation \cite{SU_3,SU_n}
\begin{equation}
W_{1}W_{0,2}=W_{0} W_{1,2}+W_{2}W_{0,1}\,,
\end{equation}
we obtain a functional relation between Wronskians of rotated
solutions: 
\begin{equation}\label{functional relation}
C^{(1)}(E,\bar{E},\mathbf{g})\,W_{1}W_{1,2}-W_{0}\,W_{1,2}-W_{2}W_{0,1} -W_{1}W_{2,3}=0\,.
\end{equation}

Alternatively, as in \eqref{Xi Psi connection A2}, we can express $\psi$ in the 
basis 
$\{\chi_{0},\chi_{1},\chi_{2}\}$ where the $\chi_j$ are solutions 
to \eqref{ODE psi x A2}
defined by the behaviour $\chi_j \sim x^{g_j}$ as $ x\to 0:$ 
\begin{equation}\label{psi in chi basis}
\psi \equiv W_{0}= Q_{0}(E, \bar{E}, \mathbf{g}) \,\chi_{0} + Q_{1}(E, \bar{E}, \mathbf{g}) \,\chi_{1} + Q_{2}(E, \bar{E}, \mathbf{g}) \,\chi_{2}\,.
\end{equation}
Using \eqref{psi in chi basis}, we will obtain from \eqref{functional relation} 
 a functional relation that is independent of $x,{\bar x}$.
First we note, from the large- and small-$x$ asymptotics of $\psi_k$
and $\chi_j$ respectively, we have
\bea \label{W shift}
W_{k,k+1,\dots,
  k{+}m}(x,\bar{x},E,\bar{E})&=&\omega^{m(3{-}m)k/2} \, 
W_{k,k+1,\dots, k{+}m}(\omega^{{-}k}x,\omega^k
\bar{x},\omega^{{-}3Mk}E,\omega^{3Mk}\bar{E}) \nn \\[2pt] 
 \chi_{j} (x,\bar{x}, E, \bar{E},\mathbf{g}) &=&  \omega
^{kg_{j}}\,  \chi_{j}( \omega^{-k}x,\omega^{k}
\bar{x},\omega^{-3kM}E,\omega^{3kM}\bar{E},\mathbf{g})~.
\eea 

We insert \eqref{psi in chi basis} into \eqref{functional relation},
then use  \eqref{W shift} to write all
Wronskians  in terms of the basic Wronskians $W_{0,1,\dots,m-1}\,.$
Since $g_{0}{<} g_{1}{<} g_{2}$, the basic 
Wronskian functions  have leading order behaviour 
\eq 
\label{psi leading order x_0}
W_{0,1,\dots,m{-}1} (x,\bar{x},E,\bar{E},\mathbf{g})
\sim Q^{(m)}(E,\bar E, {\bf g})\,
x^\alpha \,, \quad x \to 0 
\en
where $\alpha={\frac{m(1{-}m)}{2} {+} \sum_{j=0}^{m{-}1} g_j   }\,~.
$
Therefore, taking into account 
 \eqref{psi leading order x_0}, the coefficient of the
leading order term $x^{2g_0+g_1-1}$ for small $x$ of \eqref{functional relation}
 is 
\begin{multline}\label{functional relation final}
C^{(1)}(\omega^{3M}E,s) \,  Q^{(1)}(E,s) \, Q^{(2)}(E,s)  =  Q^{(1)}(\omega^{3M} E,s) \, Q^{(2)}(E,s) \, \omega^{g_{0}-1}  \\
   \hspace{-0.12cm}
{+}  Q^{(1)}(\omega^{-3M}E,s) \, Q^{(2)}(\omega^{3M} E,s) \,
   \omega^{g_{1}{-}1}  {+} Q^{(1)}(E,s) \, Q^{(2)}(\omega^{-3M}E,s) \,
   \omega^{g_2-1}\,,
\end{multline}
where $E\bar{E}=s^{6M}$. We should note that  $Q^{(1)}(E,\bar E,{\bf g})= Q_{0}(E,\bar E,{\bf g})$ and 
\eq\label{Q^2 in terms of Q^1}
Q^{(2)}(E,\bar{E},\mathbf{g}) =  \omega^{1{-}g_{1}}\,
Q_{0}(E,\bar{E},\mathbf{g})Q_{1}(\omega^{-3M}E,\omega^{3M}\bar{E},\mathbf{g}){-}
\omega^{1{-}g_{0}}\, 
Q_{0}(\omega^{-3M}E,\omega^{3M}\bar{E},\mathbf{g}) Q_{1}(E,\bar{E},\mathbf{g})\,.
\en
In terms of the  functions
\begin{equation}
T^{(1)}(E,s)= C^{(1)}(\omega^{3M}E,s)  \,, \ 
Q^{(m)}(E,s)=A^{(m)}(\omega^{-3M(m-1)/2}E,s)\,,
\end{equation}
the functional relation \eqref{functional relation final} 
corresponds to the dressed vacuum form for the 
 transfer matrix eigenvalue $T^{(1)}(E,s)$. The equivalent relation in
 the conformal case appears in (5.10) of \cite{W3}.

Now suppose that the zeros of the  functions $A^{(m)}(E)$ are at
$E_k^{(m)}$ for $k=1,2,\dots,\infty.$ Evaluating \eqref{functional
  relation final} at $E=E_k^{(m)}$ we obtain a 
set of Bethe Ansatz equations  of 
$A_2$-type for the Bethe roots $\{E_{k}^{(m)}\}$:
\bea 
\label{BAeqns}
\frac{ A^{(1)}(\omega^{3M}E_{k}^{(1)},s)
  \,A^{(2)}(\omega^{-\frac{3M}{2}}E_{k}^{(1)},s)
}{A^{(1)}(\omega^{-3M}E_{k}^{(1)},s)
  \,A^{(2)}(\omega^{\frac{3M}{2}}E_{k}^{(1)},s) } &=&
-\omega^{g_{1}-g_{0}}\,, \quad k=1,2,\dots,\infty \\
\nonumber \\
\frac{ A^{(1)}(\omega^{-\frac{3M}{2}}E_{k}^{(2)},s)
  \,A^{(2)}(\omega^{3M}E_{k}^{(2)},s)
}{A^{(1)}(\omega^{\frac{3M}{2}}E_{k}^{(2)},s)
  \,A^{(2)}(\omega^{-3M}E_{k}^{(2)},s) } &=&
-\omega^{g_{2}-g_{1}}\,,\quad k=1,2,\dots,\infty \,.
\end{eqnarray}

\resection{The $A_{n{-}1}$ cases}\label{sec: Generalised An}
We now extend the results of the previous section to all 
$A_{n{-}1}^{(1)}$ affine Toda models. 
We require an $n{-}1$-parameter family of solutions
to the Toda equations \eqref{Generalised Toda An} 
that are real, have periodicity  
$\eta_{i}\big(\rho, \phi+ 2\pi/nM\big)= \eta_{i}(\rho,\phi)\,,$
are  finite everywhere except at $\rho=0$
 and behave asymptotically as 
\bea \label{small_rho_A_n}
\eta_{i}(\rho,\phi) &=& (n-i-g_{n-i}) \ln \rho +O(1)\,,  
\quad \quad \ \rho\rightarrow 0 \,,\\ 
\label{large_rho_A_n}
\eta_{i}(\rho,\phi) &=& \fract{1}{2}(2i-1-n)M\,  \ln \rho +o(1) \,,
 \qquad  \!\! \rho \rightarrow \infty\,. 
\eea
The parameters $g_i$ are real constants satisfying $\sum_{i=0}^{n-1}g_{i}=
n(n-1)/2$ 
and, to ensure \eqref{small_rho_A_n} is the leading order
contribution for small-$\rho$, 
\begin{equation}
g_{0} < g_{1} < \cdots < g_{n-1}  \quad , \quad g_{0} + n > g_{n-1}\,.
\end{equation}

We now recast the linear problem \eqref{linear
  prob generalised An} as  two systems of $n$ first order
linear ODEs for the components of $\mathbf{\Psi} = \left( \Psi_{1},\ldots, \Psi_{n}
\right)^{\mathrm{T}}$ with solution of the form  
\bea 
\label{sol_Psi_components}
\Psi_{i}(z,\bar{z},\lambda)&=& \left\{ \begin{array}{cll} 
-\lambda^{-1} \mbox{e}^{\eta_{i}-\eta_{i+1}}
\big(\partial_{z}\Psi_{i+1} + \partial_{z} \eta_{i+1} \Psi_{i+1} \big)
& , & i=1,\ldots, n-1 \\
\mbox{e}^{\eta_{n}} \psi & ,
& i=n\,,
\end{array}\right.
\\[2pt] 
\label{sol_Psi_bar_components}
&=& \left\{ \begin{array}{cll} 
\mbox{e}^{- \eta_{1} } \bar{\psi} & \! \quad ,
& i=1\\
-\lambda \, \mbox{e}^{\eta_{i-1}-\eta_{i}}
\big(\partial_{\bar{z}}\Psi_{i-1} - \partial_{\bar{z}} \eta_{i-1}
\Psi_{i-1} \big)    &  \! \quad ,& i=1,\ldots, n-1 \,. 
\end{array}\right.
\eea
Eliminating either 
the components $\Psi_{1},\ldots, \Psi_{n-1}$  or
$\Psi_{2},\ldots,\Psi_{n}\,,$ 
we obtain
$n^{\rm th}$-order differential equations for $\psi(z,\bar z, \lambda)$ or ${\bar
  \psi}(z,\bar z, \lambda)$ respectively:
\bea 
\label{ODE_psi_An}
\Big( (-1)^{n+1} D_{n}(\eta) + \lambda^{n}p(z) \Big) \psi&=&0\,,\\ 
\label{ODE_psi_bar_An}
\Big( (-1)^{n+1} \bar{D}_{n}(\eta) + \lambda^{-n}p(\bar{z}) \Big) \bar{\psi}&=&0\,,
\eea
where the  $n$\textsuperscript{th}-order differential operators are defined by 
\begin{eqnarray}\label{D_Dbar_operators_An}
D_{n}(\eta)= \left(  \partial_{z} + 2\,\partial_{z}\eta_{1} \right) \left( \partial_{z} + 2\,\partial_{z}\eta_{2}  \right)  \cdots \left( \partial_{z} + 2\,\partial_{z}\eta_{n} \right)  \,,\\
\bar{D}_{n}(\eta)=\left(  \partial_{\bar{z}} - 2\,\partial_{\bar{z}}\eta_{n} \right) \cdots \left( \partial_{\bar{z}} - 2\,\partial_{\bar{z}}\eta_{2}  \right) \left( \partial_{\bar{z}} - 2\,\partial_{\bar{z}}\eta_{1} \right)  \,.
\end{eqnarray}

\subsection{The Q functions}

We focus on equation \eqref{ODE_psi_An} for $\psi $ and
treat  the  variable $\bar{z}$ as a parameter. 
We set  
$\mathbf{g}= \{ g_{0}, \ldots, g_{n-1} \}\,.$ 
Given the asymptotic behaviour \eqref{small_rho_A_n} of 
$\eta_{i}$, in the $\rho \rightarrow 0$ limit  the 
differential operator $D_n(\eta)$ becomes   
\begin{equation}\label{ODE_psi_An_0}
D_{n}(\mathbf{g}) = \left( \partial_{z} - \frac{ g_{n{-}1}{-}(n{-}1) }{z}
\right) 
 \left( \partial_{z} - \frac{ g_{n{-}2}-(n{-}2) }{z}
\right) 
\cdots \left( \partial_{z} - \frac{ g_{0}}{z}\right)\,.
\end{equation}
Hence we read off from \eqref{ODE_psi_An_0} that the 
solutions to  \eqref{ODE_psi_An} behave as $z^{g_{j}}$ as $z \to 0\,.$
Therefore the linear problem \eqref{linear prob generalised An} has 
$n$ solutions defined as $\rho \to 0$ by  
\begin{equation}\label{Psi_components_0}
\mathbf{\Xi}_{j} \sim ( \underbrace{  0, \ldots,0,
  \mbox{e}^{(g_{j}-j)(\theta +i \phi)},0 \ldots, 0 }_{n
  \;\mathrm{components}}  )^{\mathrm{T}}\,,
\end{equation}
where the  $(n{-}j)^{\rm th}$ component is non-zero. 
The  solutions $\mathbf{\Xi}_{j}$  respect the symmetries
of the linear 
problem induced by the transformations  $\widehat{\Omega}$,
$\widehat{S}$: 
\begin{equation}
\mathbf{\Xi}_{j} \left(  \rho, \phi + \fract{2\pi}{nM}, \theta - \fract{2 \pi i}{nM}  \right) = \mathbf{\Xi}_{j} \left( \rho, \phi, \theta \right)
\end{equation}
and 
\begin{equation}
S \, \mathbf{\Xi}_{j} \left( \rho, \phi, \theta - \fract{2 \pi i}{n}  \right) = \exp \left(  -g_{j} \fract{2 \pi i}{n} \right) \mathbf{\Xi}_{j} \left( \rho, \phi, \theta \right) \,.
\end{equation}
Linear independence of the set $\{ \mathbf{\Xi}_0,\dots,
  \mathbf{\Xi}_{n{-}1}\}$ follows from    $\det \left( \mathbf{\Xi}_{0}, \ldots,
  \mathbf{\Xi}_{n{-}1} \right) = -1\,.$

On the other hand, in the large-$\rho$ limit there exists a  solution
to the ODE 
\eqref{ODE_psi_An} with asymptotic representation for $M>1/(n{-}1)$
given by
\begin{equation}\label{WKB psi z and zbar final An}
 \psi \sim z^{-(n-1)M/2} \exp \left( -\lambda\, \frac{z^{M+1}}{M+1} -
   \lambda^{-1} \frac{\bar{z}^{M+1}}{M+1} \right)\,, \quad\quad \rho \to \infty\,.
\end{equation}
This is the unique solution that decays fastest in the sector of the
complex plane defined by 
$z=\rho\, \exp(i\phi)$ with $|\phi|<(n+1)\pi/n(M+1)\,.$ 
Therefore in the limit $\rho \rightarrow \infty$ a solution to the linear problem \eqref{linear prob generalised An} reads
\begin{equation}\label{sol Psi large rho An}
\mathbf{\Psi}(\rho,\phi,\theta,\mathbf{g}) \sim (\Psi_{1}, \ldots,
\Psi_{n})^{{\rm T}} \, \exp \left( -2 \, \frac{\rho^{M+1}}{M+1} \cosh (\theta + i\phi(M+1)) \right)\,,
\end{equation}
where
\begin{equation}\label{components sol Psi large rho}
\Psi_{j}= \exp \left(  i \phi \, \frac{n-(2j-1)}{2}\,M \right)\,.
\end{equation}
Expanding $\mathbf{\Psi}$ in the basis of solutions
$\{\mathbf{\Xi}_{0} ,\mathbf{\Xi}_{1},\dots,\mathbf{\Xi}_{n-1}\}$
yields 
\begin{equation}\label{Psi in Xi basis An}
\mathbf{\Psi} (\rho, \phi, \theta, \mathbf{g}) = \sum_{j=0}^{n-1} Q_{j} (\theta, \mathbf{g}) \, \mathbf{\Xi}_{j} (\rho, \phi, \theta, \mathbf{g})\,.
\end{equation}

\subsection{Conformal limit}\label{sec:Conformal limit An}

We now check that in the massless limit described below the differential
equations \eqref{ODE_psi_An} and \eqref{ODE_psi_bar_An} are consistent
with the $n^{\rm th}$-order differential equations of 
the relevant conformal quantum integrable models~\cite{Suzuki_A_n,SU_n}.
It is convenient to  define
\begin{equation}\label{change of vars An}
x=z\,\mbox{e}^{\frac{\theta}{M+1}}\,, \;  \bar{x} = \bar{z}\,
\mbox{e}^{-\frac{\theta}{M+1} }\,, \; E= s^{nM} \mbox{e}^{
  \frac{n\theta M}{M+1} }\,, \; \bar{E} = s^{nM} \mbox{e}^{
  -\frac{n\theta M}{M+1} }\,. 
\end{equation}
The massless limit of the ODE 
\eqref{ODE_psi_An} in terms of  $\psi$  is obtained by first taking
$\bar{z}\rightarrow 0$
with $z$ finite and small, then taking the limit $z\sim s \to 0$ while 
$\theta \rightarrow +\infty\,.$ This process yields 
\begin{equation}\label{ODE psi new vars An 2}
\Big( (-1)^{n+1} D_{n}(\mathbf{g}) + p(x,E) \Big) \psi(x,E)=0\,, \quad p(x,E) = x^{nM} - E \,,
\end{equation}
where the operator $D_n({\bf g})$ defined in \eqref{ODE_psi_An_0} is
now a function of $x$. This is precisely   the $n$\textsuperscript{th}-order ODE 
appearing in the massless
$SU(n)$ ODE/IM correspondence \cite{Suzuki_A_n, SU_n}.

Similarly,  sending the parameter $z \to 0$ 
in the ODE for $\bar{\psi}$ \eqref{ODE_psi_bar_An}  while keeping $\bar z$ small but finite, then 
 taking the limit $\bar z \sim s \to 0 $ as  $\theta \rightarrow
 -\infty$ we find 
\begin{equation}\label{ODE psi bar new vars An 2}
\Big( (-1)^{n+1} D_{n}(\mathbf{g}^{\dagger}) + p(\bar x,\bar E) \Big)
\bar{\psi}(\bar x , \bar E) =0\,, 
\end{equation}   
 with ${\bf g}^\dag {=} ( g_0^\dag,g_1^\dag,\dots,g_{n-1}^\dag)$
and $g_j^\dagger = n{-}1{-}g_{n-1-j}$. Equation \eqref{ODE
  psi bar new vars An 2} is the adjoint equation to \eqref{ODE psi new
  vars An 2}.  This equation also appears naturally 
in the 
$SU(n)$ ODE/IM correspondence \cite{SU_n}.

\subsection{Functional relations}\label{sec:functional relations
  A_n}
By establishing a set of functional relations satisfied by functions
of  the $Q_j(\theta,{\bf
  g}),$  we will  obtain the $A_{n{-}1}$
Bethe ansatz systems. 
Using the  variables \eqref{change of vars An},  
we define  solutions to
\eqref{ODE_psi_An} for integer $k$  by 
\begin{equation}\label{psi k An} 
\psi_{k}(x,\bar{x},E, \bar{E}, \mathbf{g}) = \omega^{(n-1)k/2} \psi
(\omega^{-k}x,\omega^{k}\bar{x},\omega^{-nkM} E,\omega ^{nkM} \bar{E},
\mathbf{g} ) \,, \ \quad 
\omega = \mbox{e}^{\frac{2\pi i}{n(M+1)}}\,
\end{equation}
with large-$x$ asymptotics  for fixed real $\bar{x}$
\begin{equation}\label{WKB psi k An}
\psi_{k} \sim \omega^{k(n-1)(M+1)/2} \frac{x^{-M}}{i^{(n-1)/2}\sqrt{n}} \exp \left( -\omega^{-k(M+1)}\frac{x^{M+1}}{M+1} -\omega^{k(M+1)} \frac{\bar{x}^{M+1}}{M+1} \right)\,.
\end{equation}
These asymptotics are valid for  $x \in
\mathcal{S}_{k-\frac{n}{2}} \cup \dots  \cup
\mathcal{S}_{k+ \frac{n}{2}}$ where  $$
\mathcal{S}_{k}:\left|\arg x-\frac{2k\pi}{n(M+1)}\right|<\frac{\pi}{n(M+1)}\,.
$$ 
Moreover,  each $\psi_k$ is the unique solution that decays to zero
fastest for large $x$ within the Stokes sector ${\cal S}_k$. By
construction $W[\psi_{1},\dots,\psi_{n}]=1$, so we have a 
basis of solutions to 
\eqref{ODE_psi_An}  and may  write 
\begin{equation}\label{psi0 in terms of psi12n}
\psi = \sum_{k=1}^n C^{(k)}(E,\bar{E},\mathbf{g})\, \psi_{k} \,.
\end{equation}
Since the next steps follow \cite{SU_n} closely  we omit some of the details. 
Using the determinant 
relations given in \cite{SU_n}, the 
general-$n$ version of the functional relation \eqref{functional
  relation} 
 is  
\eq \label{cw}
C^{(1)}(E,\bar E , {\bf g}) \prod_{j=0}^n W_1^{(j)} =\sum_{m=0}^{n-1} 
\lf (\prod_{j=0}^{m-1} W_1^{(j)} \ri) W_2^{(m)} W_0^{(m+1)} 
\lf (\prod_{j=m+2}^{n} W_1^{(j)} \ri)\,,
\en
where 
\eq W_k^{(m)} = W_{k,k+1,\dots,k+m}(x,\bar x , E,\bar E ,{\bf g})\,.\en
The elimination of the dependence of \eqref{cw} on $x,\bar
x$  is achieved by expanding \eqref{cw} in the
alternative basis  given by solutions to \eqref{ODE_psi_An} that
have 
small-$x$ behaviour defined by the components
of ${\mathbf \Xi}_j$ \eqref{Psi_components_0}. 
 The Wronskians $W_k^{(m)}$ may be written explicitly in terms of this
 basis (see (5.5) of \cite{SU_n}) using the bottom component of the
 expansion \eqref{Psi in Xi basis An} of ${\bf \Psi}$. Let
 $Q^{(m)}(\omega^{-nMk }E,\omega^{-nMk } \bar E ,\bg) $ denote the 
 coefficient of the dominant term of  $W_k^{(m)}$ as $x\to 0\,.$ 
In particular, from 
\eqref{Psi in Xi basis An} we have $Q_{0}=Q^{(1)}\,.$ 
Inserting the resulting expansions of $W_k^{(m)}$ into \eqref{cw}, we
find  the  coefficient of the 
leading order term of \eqref{cw} in the limit as $x \to 0$ reads: 
\bea 
&& \hspace{-0.8cm} C^{(1)}(\omega^{nM}E,s ) \prod_{j=0}^n Q^{(j)}(E,s) =\nn \\ 
&&\hspace{-0.65cm} \sum_{m=0}^{n-1}  
\lf (\prod_{j=0}^{m-1} Q^{(j)}(E,s) \ri) 
\omega^{\beta_{m{+}1}{-}\beta_m} Q^{(m)}(\omega^{-nM}E, s)
 Q^{(m{+}1)}(\omega^{nM}E, s) 
\lf (\prod_{j=m{+}2}^{n} Q^{(j)}(E,s) \ri)  \label{tq}
\eea
where the dependence of all functions on $\bg$  has been omitted, 
 $\beta_{m} = \sum_{j=0}^{m-1}g_{j} -
m(n-1)/2$ and $E\bar E=s^{2nM}\,.$    The functional relation \eqref{tq} is expected
to coincide with the dressed vacuum form for one of the  transfer
matrix eigenvalues of the 
corresponding massive quantum integrable model.  Related functional
equations derived for the  ${\cal W}_N$ conformal field
theory appear in \cite{wn}.  

We make one more redefinition:   
 $A^{(m)}(\omega^{-nM(n-1)/2}E, s)= Q^{(m)}(E,s)\,.$  
Finally, setting $E_{k}^{(m)}$ to denote  a zero of $A^{(m)}(E,s)$ we
  obtain from \eqref{tq} the $A_{n-1}$ Bethe ansatz equations: 
\begin{equation}\label{BA eqns SU_n}
\frac{A^{(m-1)}(\omega^{-nM/2} E_{k}^{(m)},s)}{A^{(m-1)}(\omega^{nM/2} E_{k}^{(m)},s)} \, \frac{A^{(m)}(\omega^{nM} E_{k}^{(m)},s)}{A^{(m)}(\omega^{-nM} E_{k}^{(m)},s)} \, \frac{A^{(m+1)}(\omega^{-nM/2} E_{k}^{(m)},s)}{A^{(m+1)}(\omega^{nM/2} E_{k}^{(m)},s)} = -\omega^{-2 \beta_{m} + \beta_{m-1} + \beta_{m+1}}\,.
\end{equation}

\section{Conclusions}

In this paper we have demonstrated how the massive generalisation of the
$A_{n}$ ODE/IM correspondence of \cite{SU_n} can be constructed
starting from the system of classical partial differential equations
appearing in   $A_{n}^{(1)}$ Toda field theory. Moreover, we have obtained 
functional relations satisfied by the massive $Q$ functions 
and  derived the corresponding Bethe ansatz systems.

Of immediate interest is to further study the analytic properties of
the $T$ and $Q$ functions obtained in sections~\ref{sec: Generalised A_2}
and~\ref{sec: Generalised An}, and fully explore the integrable features.
The $Q$ functions 
considered here correspond to the vacuum eigenvalues of the
corresponding quantum integrable model. It would be interesting to
develop to all $A_n$ models the recent work  of Bazhanov and Lukyanov
\cite{Bazhanov:2013cua}, which extends the massive sine Gordon/sinh
Gordon  correspondence  to higher-level eigenvalues of the quantum integrable
model.

\medskip
\noindent{\bf Acknowledgments --} 
We would like to thank P.A. Clarkson, A.V. Mikhailov,
G. Papamikos and J.P.  Wang 
for useful discussions. 
This work was supported by the Engineering and Physical Sciences
Research Council [grant number EP/G039526/1].

\medskip
\noindent{\bf Note added --} This paper is based on chapters 5 and 6
of \cite{pma}, submitted by PA for the award of PhD in October 2013.  As
we were completing the article,   
we became aware of the preprint \cite{Ito:2013aea}, which also reports results 
on the $A_{n}^{(1)}$ massive ODE/IM correspondence and extends the
analysis to all
simply laced affine Lie algebras and $A_{2n}^{(2)},D_4^{(3)}$ and $G_2^{(1)}$.

\end{document}